\documentclass[12pt]{article}

\usepackage{amsmath,amsfonts}
\makeatletter \@addtoreset{equation}{section}

\usepackage{cite}
\usepackage{bm}
\usepackage{dcolumn}
\usepackage[pdftex]{graphicx}
\usepackage{wrapfig}
\usepackage{sidecap}
\usepackage{graphicx}
\usepackage{caption}
\usepackage{subcaption}
\usepackage{graphicx}
\usepackage{dcolumn}
\usepackage{eufrak}
\usepackage{yfonts}
\usepackage{dsfont}
\usepackage{bbold}
\DeclareMathAlphabet{\mathpzc}{OT1}{pzc}{m}{it}
\makeatother
\def\one{{\hbox{ 1\kern-.8mm l}}}
\newcommand{\Dslash}{\not{\hbox{\kern-4pt $D$}}}
\newcommand{\pdslash}{\not{\hbox{\kern-2pt $\partial$}}}
\newcommand{\be}{\begin{equation}}
\newcommand{\bea}{\begin{eqnarray}}
\newcommand{\eea}{\end{eqnarray}}
\newcommand{\ba}{\begin{array}}
\newcommand{\ea}{\end{array}}
\newcommand{\ee}{\end{equation}}

\begin{document}

\begin{titlepage}
\vspace*{1mm}%
\hfill%

\vspace*{15mm}%
\begin{center}

{{\Large {\bf Fractional Galilean Symmetries}}}

\vspace*{15mm} \vspace*{1mm} { Ali Hosseiny$^{a,c}$, and Shahin Rouhani$^{b,c}$}

\vspace*{.4cm}

{\it ${}^a$ Department of Physics, Shahid Beheshti University \\
G.C., Evin, Tehran 19839, Iran. }

\vspace*{.4cm}

{\it ${}^b$ Department of Physics, Sharif University of Technology \\
P.O. Box 11165-9161, Tehran, Iran. }

\vspace*{.4cm}

{\it ${}^c$ School of Particles and Accelerators, \\ Institute for Research in Fundamental Sciences (IPM)\\
P.O. Box 19395-5531, Tehran, Iran. \\}

\vspace*{2cm}

e-mail: {~~ \tt al$\_$hosseiny@sbu.ac.ir} ~~and {\tt rouhani@ipm.ir}

\end{center}

\begin{abstract}
We generalize the differential representation of the operators of the Galilean algebras to include fractional derivatives. As a result a whole new class of scale invariant Galilean algebras are obtained. The first member of this class has dynamical index $z=2$ similar to the Schr\"odinger algebra. The second member of the class has dynamical index $z=3/2$, which happens to be the dynamical index Kardar-Parisi-Zhang equation.
\end{abstract}

\end{titlepage}




\section{Introduction}

Locality has been the guiding principal of many branches of physics. This has come about
in opposition to the concept of action at a distance. That is to say for something to influence
another thing an entity such as a field or particle must travel between the two points. 
However phenomena such as the EPR experiments point to the plausibility of
non-local theories in quantum mechanics \cite{Giustina}. Another setting in which non-local theories arise is equilibrium thermodynamics, where systems have had enough time to reach a state of equilibrium where long range correlations have come into existence. An example of recent work in this direction is given by \cite{Paulos}, where conformal invariance of long range Ising model is discussed. This means that non-local theories may be acceptable.  If locality is given up a very rich arena of possible models in theoretical physics arises, one of which is the option of having fractional derivatives. Consequently this opens up the field for seeking symmetry algebras which are composed of derivatives with fractional order. In this paper we are going to see what we can get if Galilean symmetry is augmented with fractional derivative operators. We find a whole new class of scale and Galilean invariant symmetries, with previously unknown dynamic exponents. 
Galilean symmetries have received some attention recently due to their appearance in the AdS/CFT correspondence \cite{Balasubramanian}-\cite{Fareghbal}. The Galilean symmetries are composed of translations in time and space plus rotations in space and boosts to inertial frames that move with constant velocity. One can add to this set, scale transformations and boosts to frames of reference which conserve higher derivatives such as accelerations plus “toroidal” transformations of the time component. These symmetries lead to L-Galilei algebras \cite{{Henkelprl},{Negro}} where L is an integer, for related works see \cite{hosseinycentral}-\cite{Galajinsky}. The disappointment in L-Galilean symmetries is that not all dynamical exponents “$z$” arise, although the first two members of this class i.e. Galilean Conformal Algebra (GCA) $z=1$ \cite{Havas}-\cite{Grumiller} and Schr\"odinger symmetry $z=2$, \cite{Niederer}-\cite{Fushchych} are well known. Attempts to obtain other dynamical exponents exist \cite{Henkelpheno} where the dynamical index of $z=\frac{3}{2}$ was obtained. However this remained a singular case. Other cases have been suggested by Henkel and Stoimenov in which further values of dynamical index are obtained using the age subalgebra of the Schr\"odinger algebra \cite{{henkelstoia},{henkelstoib}}. However the algebra only closes on the solution space. Besides the exponent, the geometrical meaning of infinitesimal fractional algebra has been analyzed in the context of the Schr\"odinger algebra in \cite{henkelstoib}.

In this paper we show that admitting fractional derivatives allows a new class of Galilean symmetries, which we call F-Galilean where F is an integer and the dynamical exponent is given by 
\bea
z=\frac{F+1}{F}
\eea
Thus for $F=1$, we have the well-know exponent $z=2$ (This symmetry is smaller than Schr\"odinger symmetry) and for $F=2$ we get $z=\frac{3}{2}$, i.e. the well-known exponent of the Kardar-Parisi-Zhang (KPZ) equation \cite{Kardar}.  The price we paid for this extension is non-locality through introduction of fractional derivatives. 
This paper is organized as follows, in section 2 we give a brief introduction to L-Galilei algebras, which in fact encompasses all known galilean algebras. In Section 3 we introduce fractional derivatives and some of their properties. In section 4 we derive the new class of galilean algebras which posses differential operator representations including fractional derivatives. It is in fact through the consistency of these representations that we derive this new class of non-local galilean symmetries.


\section{L-Galilei Algebra}

L-Galilei algebra were introduced independently by Henkel \cite{Henkelprl} and Negro, del Olmo and  Rodr\'{\i}guez-Marco \cite{Negro}. Here we briefly review their method. First, let us recall the two well-known non-relativistic symmetries: Schr\"odinger and conformal Galilean algebras.\\\\

{\bf -Schr\"odinger Algebra }\\

Schr\"odinger algebra consists of the following generators. The  center is formed by the generators of the Galilean transformations
\bea
P=-\partial_x,\;\;\;\;\;\;\;\;\;\;\; H=-\partial_t,\;\;\;\;\;\;\;\;\;\;B=-t\partial_x
\eea
which guarantees invariance under translations in space and time, plus Galilean boost. 

Besides the Galilean generators, the Schr\"odinger algebra has three more generators. Two of them are, the dilation generator:
\bea
D=-(t\partial_t+\frac{1}{2} x\partial_x),
\eea
along with the special Schr\"odinger generator
\bea
C=-(t^2\partial_t+tx\partial_x).
\eea
The last generator is the mass operator which is the outcome of commutators of $B$ and $P$
\bea
[B,P]=M_0.
\eea
The interesting point with Schr\"odinger algebra is that it has an $SL(2,R)$ subalgebra: $H,\; D$ and $C$. Commutators read as
\bea
[D,C]=-C\;\;\;\;\;\;\;\;\; [D,H]=H\;\;\;\;\;\;\;\;\;[C,H]=2D.
\eea
In this Cartan subalgebra $H$ produces translation in time and $D$ produces scale transformations
\bea
x\rightarrow\lambda x,\;\;\;\;\;\;\;\;\;\;\;\;\;\;\;\;t\rightarrow\lambda^2 t.
\eea
Together with $C$, the subalgebra guarantees global conformal transformation in temporal direction
\bea
t\rightarrow\frac{\alpha t+\beta}{\gamma t + \delta}\;\;\;\;\;\;\;\;\;\alpha\delta-\beta\gamma=1
\eea
along with its spatial counterpart
\bea
 x\rightarrow\frac{ x}{(\gamma t+\delta)}.
\eea
Adding Galilean boost to these elements we end up with the following transformation in space and time
\bea\begin{split}
t\rightarrow\frac{\alpha t+\beta}{\gamma t + \delta}\;\;\;\;\;\;\;\;\;\;\;
x\rightarrow\frac{x+c_1t+c_0}{\gamma t+\delta} \;\;\;\;\; \;\;\;\;\cr \;\;\;\;\;\;\;\;where\;\;\;\;\;\;\;
\alpha\delta-\beta\gamma=1
\end{split}\eea\\
Extension to other dimensions is straightforward. Actually the algebra reads
\bea\begin{split}
&P_i=-\partial_i\;\;\;\;\;\;\;\;\;\;\;\;\;\;\;\;\;\;B_i=-t\partial_i\;\;\;\;\;\;\;\;\;\;\;\;\;\;\;\;\;\;\;\;J_{ij}=x_j\partial_i-x_i\partial_j
\cr& H=-\partial_t\;\;\;\;\;\;\;\;\;\;\;\;\;\;D=-t\partial_t-\frac{1}{2}x_i\partial_i\;\;\;\;\;\;\;\;\;\;\;\;C=-(t^2\partial_t+tx\partial_x).
\end{split}\eea\\
While $P_i$ and $B_i$ are vectors, $H, D$, and $C$ commute with $J_{ij}$.\\

{\bf -Conformal Galilean Algebra}\\

The other well-known non-relativistic algebra; the Conformal Galilean Algebra (CGA) \cite{Havas}; similar to the Schr\"odinger algebra consists of six generators. Similar to the Schr\"odinger algebra, CGA contains a $SL(2,R)$ subalgebra 
\bea
H=-\partial_t,\;\;\;\;\;\;\;\;\;\;\;\;\;\;D=-t\partial_t-x\partial_x\;\;\;\;\;\;\;\;\;\;\;\;\;\;C=-t^2\partial_t-tx\partial_x
\eea
Unlike the Schr\"odinger algebra, CGA, dilation generator scales space and time isotropically
\bea
x\rightarrow\lambda x,\;\;\;\;\;\;\;\;\;\;\;\;\;\;\;\;t\rightarrow\lambda t.
\eea
 \\
{\bf - L-Galilei Algebra}\\

As it was seen, both Schr\"odinger algebra and CGA contain a $SL(2,R)$ subalgebra which guarantees global transformation in temporal direction. In other words we have translation in time $H=\partial_t$. As well we have scaling in space and time  $D=-(t\partial_t+\frac{1}{z}x\partial_x)$ which generally produces an anisotropic dilation in space and time
\bea
x\rightarrow\lambda x,\;\;\;\;\;\;\;\;\;\;\;\;\;\;\;\;t\rightarrow\lambda^z t.
\eea
To close $SL(2,R)$ subalgebra we have no choice other than to define $C$ and it reads as
\bea
C=-(t^2\partial_t+\frac{2}{z}tx\partial_x).
\eea
To build L-Galilei algebra we start with this $SL(2,R)$ subalgebra we add two generators of transformation in space $P=-\partial_x$ and the Galilean boost $B=-t\partial_x$. Now, we observe that the action of $D$ counts the scaling weights of all generators
\bea
[D,P]=\frac{1}{z}P\;\;\;\;\;\;\;\;\;\;\;[D,B]=(\frac{1}{z}-1)B.
\eea
On the other hand C raises scaling weights of the generators.
\bea
[C,P]=(\frac{2}{z}-1)B\;\;\;\;\;\;\;\;\;\;\;[C,B]=(\frac{2}{z}-1)tB.
\eea
Note that the action of $C$ has produced a new operator $tB$ which has a higher weight than $B$. So, we define the new generators $P_q=-t^q\partial_{x}$, such that for $q=0$ it represents momentum and for $q=1$ it represents boost. Now, C raises the value of $q$ and we can simply write
\bea
[C,P_q]=(\frac{2}{z}-q)P_{q+1}
\eea
If we desire the algebra to close for some finite value $L$ then we set $2/z-L=0$. So the algebra closes if $z=\frac{2}{L}$ where $L$ is an integer.
Gathering all commutation relations we have
\bea \label{lgalilei} \begin{split}
&[D,H]=H,\;\;\;\;\;\;\;\;\;\;\;\;\;\;\;\;\;\;\;\;\;\;\;\;\;[D,C]=-C,\;\;\;\;\;\;\;\;\;\;\;\;\;\;\;\;\;\;\;\;\;\;\;\;[C,H]=2D,\cr&[H,P_q]=-qP_{q-1}\;\;\;\;\;\;\;\;\;\;\;\;\;\;\;\;\;[D,P_q]=(l-q)P_q,\cr&[C,P_q]=(2l-q)P_{q+1}
\end{split}\eea 
The obtained L-algebra is the infinitesimal form of the following transformations
\bea \label{} \begin{split}
&\vec x\rightarrow\frac{x+t^{2l}\vec c_{2l}+...+t\vec c_1+\vec c_0}{(\gamma t+\delta)^{2l}}\;\;\;\;\;\;\;\;\;\;\;\;\;\;\;\;\;\;\;\;t\rightarrow\frac{\alpha t+\beta}{\gamma t + \delta}
\end{split}\eea 
where ${\vec c_n\in\mathbb{R}^n}$.

The smallest element of the class is given by $L=1$ which is the Schr\"odinger algebra, widely studied in the literature. The second element is given by $L=2$ and is nothing but CGA.

 Note that although the $SL(2,R)$ generators exist in both CGA and Schr\"odinger algebra their presence is not crucial to non-relativistic algebras. In the aging problem for instance, time translation invariance is not satisfied, see \cite{Henkelp}-\cite{Henkelkpz}. For a much more comprehensive study see \cite{Henkelpheno}.
With this motivation in mind we note that by extending to fractional derivatives the algebra obtained in this section can be generalized.

\section{Fractional Derivatives}

The idea is to extend the concept of derivative $D^n=(\frac{d}{dx})^n$to non-integer values $D^a$  for all values
 of a. Our motivation behind this extension is that new symmetry groups can arise which 
cannot exist for integer values of $a$. It is expected that the local property of the differential operator will be lost when $a$ is no longer an integer. An obvious application is offered by the Brownian motion. Whilst the normal Brownian motion with an index $\alpha=1$ is related to the diffusion equation,      
\bea
<\Delta x^2>\; \sim t^{\alpha}
\eea
the sub (super) ordinate Brownian motion is described by non-integer $\alpha<1\;(\alpha>1)$ anomalous diffusion equation \cite{Metzler}  
\bea
\frac{\partial\psi}{\partial t}=\Delta^{\frac{1}{\alpha}}\psi
\eea
Usually extensions of properties of regular derivatives to the fractional ones are not straightforward. Here we are concerned with commutation relations. In regular derivatives we always have the properties of commutation between derivatives $D^m$ and $D^n$. In a sense that
\bea\label{exponent}
D^mD^nf(x)=D^nD^mf(x)=D^{m+n}f(x).
\eea
where $m , \;n \in\mathbb{N} $. Such property however is not extendable to the fractional derivatives. Actually there are functions with special singularities which prevent us from such extension. So, following \cite{Henkelpheno} we consider a space in which one can extend the law of exponents in Eq. (\ref{exponent}) to the fractional derivatives.  

Let's consider a set of numbers $e$ which is called $E-set$ where the numbers are real but not negative integer, i.e. $e\neq -(n+1)$, $n\in\mathbb N$. Then consider a (possibly infinite) positive  real interval $I$. Now, one can consider a space of functions which is called ${\cal{M}}-space$ related to the E-set
\bea
{\cal M}:={\cal M}_E(I,\mathbb{R})=\{f: I\subset\mathbb{R}\rightarrow\mathbb{R}| f(x)=\Sigma_{e\in E}f_ex^e+\Sigma_{n=0}^\infty F_n\delta^{(n)}(x); f_e, F_n\in\mathbb{R}\}
\eea
where $\delta^{(n)}(x)$ indicates the $n_{th}$ derivative of the Dirac delta function. The first set of constant terms or ($f_e$) is called the regular part of $f$ and the second set ($F_e$) is called the singular part of $f$. Now, we generalize conditions and consider a set of numbers $E=\alpha\mathbb{N}+\beta$ with $\alpha> 0$ and $\beta\neq-(\alpha(n+1)+m+1)$ where $n,m\in\mathbb{N}$. Now, ${\cal M}_E$ is the space of the functions with the form of $x^{\beta}f(x^{\alpha})$ where $f(x)$ is analytic.

For a real number $a$ we can consider a set $E^{\prime}=\{e^{\prime}|e^{\prime}=e-a; e\in E\}$ and the related space ${\cal M}^{\prime}={\cal M}_{E^{\prime}}$. Now, an operator $\partial^a:{\cal M}\rightarrow{\cal M^{\prime}}$ is considered a derivative of order $a$ if we have
\bea\begin{split}
&i)\;\;\;\partial^a(\alpha f(x)+\beta g(x))=\alpha\partial^af(x)+\beta\partial^ag(x)\;\;\;\forall \alpha, \beta \in\mathbb{R} \;and\; \;\forall f, g \in {\cal M}
\cr&ii)\;\;\;\partial^ar^e=\frac{\Gamma(e+1)}{\Gamma(e-a+1)}x^{e-a}+\Sigma_{n=0}^{\infty}\delta_{a,e+n+1}\Gamma(e+1)\delta^{(n)}(x)
\cr&iii)\;\;\;\partial^a\delta^{(n)}(x)=\frac{x^{-1-n-1}}{\Gamma(-a-n)}+\Sigma_{m=0}^{\infty}\delta_{a,m}\delta^{(n+m)}(x)
\end{split}
\eea
where $\Gamma(x)$ is the Gamma function. For such derivatives acting on ${\cal M}$ space it has been proven in \cite{Henkelpheno} that we can have the following properties
\bea\begin{split}
&\partial^a\partial^bf(x)=\partial^b\partial^af(x)=\partial^{a+b}f(x)
\cr&[\partial^a,x]f(x)=a\partial^{a-1}f(x).
\end{split}\eea
For such space where we can have the above-mentioned commutation relation we will look for some sort of fractional Galilean symmetries.


\section{Fractional Galilei Algebras}

The aim is to find a symmetry which respects Galilean transformations as well as scale
 invariance. We start with Dilation and translation generators
\bea
D=-(t\partial_t+\frac{1}{z} x\partial_x),\;\;\;\;\;\;\;\;\;\;H=-\partial_t, \;\;\;\;\;\;\;\;\;\;P=-\partial_x
\eea
which satisfies the commutation relations
\bea
[D,H]=H,\;\;\;\;\;\;\;[D,P]=\frac{1}{z}P,\;\;\;\;\;\;\;[H,P]=0.
\eea
Following \cite{Henkelpheno} we define the boost generator as
\bea
B=-t\partial_x-{\cal{M}}(\partial_x)x
\eea
To go further we better write generators in a way that their weight is represented. So, we write
\bea
X_0=D\;\;\;\;\;\;X_{-1}=H\;\;\;\;\;\;Y_+=B\;\;\;\;\;\;Y_-=P,
\eea
Now the scaling weights of generators is indicated and for the commutation relations we simply have
\bea
[X_0,X_{-1}]=X_{-1}\;\;\;\;\;[X_0,Y_-]=\frac{1}{z}Y_-.
\eea
For $Y_+$ we then expect
\bea
[X_0,Y_+]=-\frac{z-1}{z}Y_+.
\eea
Plugging in their differential forms we obtain
\bea
\frac{z-1}{z}t\partial_x-\frac{1}{z}{\cal{M}}^{\prime}(\partial_x)x\partial_x+\frac{1}{z}x{\cal{M}}(\partial_x)=\frac{z-1}{z}[t\partial_x+{\cal{M}}(\partial_x)x],
\eea
which results in
\bea
(z-2){\cal{M}}(\partial_x)=-{\cal{M}}^{\prime}(\partial_x)\partial x.
\eea
The solution is
\bea
{\cal{M}}(\partial_x)=M_0\partial_x^{2-z},
\eea
in which $M_0$ is a constant number. So, the final form of $Y_+$ is:
\bea
Y_+=-t\partial_x-M_0\partial_x^{2-z}x
\eea
Now, we go through the commutators and seek the value of $z$ for which the algebra closes. Start with
\bea
[X_{-1},Y_-]=0,\;\;\;\;\;\;\;\;[X_{-1},Y_+]=Y_-.
\eea
and observe that
\bea
[Y_+,Y_-]=-M_0\partial_x^{2-z}
\eea
We call this new operator $Q_1$. Let's investigate the commutation relation of $Q_1$ with other operators
\bea\begin{split}
&[X_0,Q_1]=\frac{2-z}{z}Q_1\;\;\;\;\;\;\;\;\;\;\;\;\;\;\;\;\;\;\;\;[X_{-1},Q_1]=0
\cr&[Y_+,Q_1]=(2-z)M_0^2\partial_x^{3-2z}\;\;\;\;\;\;\;\;\;[Y_-,Q_1]=0.
\end{split}\eea
Now we define the sequence of generators
\bea
Q_i=-M_0^i[\Pi_{j=1}^i(j-(j-1)z)]\partial_x^{i+1-iz}.
\eea
Commutators of existing generators with $Q_i$ read as
\bea\begin{split}
&[X_0,Q_i]=\frac{i+1-iz}{z}Q_i\;\;\;\;\;\;\;\;\;\;\;\;\;[X_{-1},Q_i]=0\;\;\;\;\;\;\;\;\;\;\;\;\;\;\;\;\;\;\;\;[Q_i,Q_j]=0
\cr&[Y_+,Q_i]=Q_{i+1}\;\;\;\;\;\;\;\;\;\;\;\;\;\;\;\;\;\;\;\;\;\;\;\;\;[Y_-,Q_i]=0.
\end{split}\eea
The scaling weight of $Q_i$ is $\frac{i+1-iz}{z}$ and $Y_+$ acts as ladder operator. Now, suppose that given the value of $z$, there is a number $F$ such that the scaling weight of $Q_F$ is zero. This causes the algebra to close since the ladder operator does not generate a new generator. This leads to the value
\bea
z=\frac{F+1}{F}.
\eea   
If we look at definition of $Q_i$ then we see that for such values of $F$ and $z$ $Q_F$ is a number and then the algebra will be closed. Summing up for any value of $F$ we have an algebra as follows
\bea\begin{split}
&X_0=-(t\partial_t+\frac{F}{F+1} x\partial_x)\;\;\;\;\;\;\;\;\;\;\;\;\;\;\;\;\;\;\;\;\;X_{-1}=-\partial_t
\cr&Y_+=-t\partial_x-M_0\partial_x^{\frac{F-1}{F}}x\;\;\;\;\;\;\;\;\;\;\;\;\;\;\;\;\;\;\;\;\;\;\;\;Y_-=-\partial_x
\cr&Q_i=-\frac{M_0^i}{F^i}[\Pi_{j=1}^i(F-j+1)]\partial_x^{\frac{F-i}{F}}\;\;\;\;\;\;i=1:F
\end{split}\eea
We can gather all commutators   
\bea\begin{split}
&[X_0,X_{-1}]=X_{-1}\;\;\;\;\;\;\;\;\;\;\;\;[X_0,Y_-]=\frac{F}{F+1}Y_-\;\;\;\;\;\;\;\;\;[X_0,Y_+]=-\frac{1}{F+1}Y_+\cr&[X_0,Q_i]=\frac{F-i}{F+1}Q_i\;\;\;\;\;\;\;[X_{-1},Y_+]=Y_-\;\;\;\;\;\;\;\;\;\;\;\;\;\;\;\;\;\;[X_-,Y_-]=0
\cr&[X_{-1},Q_i]=0\;\;\;\;\;\;\;\;\;\;\;\;\;\;\;\;\;[Y_+,Y_-]=Q_1\;\;\;\;\;\;\;\;\;\;\;\;\;\;\;\;\;\;\;[Y_+,Q_i]=Q_{i+1} \cr&[Y_-,Q_i]=0\;\;\;\;\;\;\;\;\;\;\;\;\;\;\;\;\;\;\;[Q_i,Q_j]=0
\end{split}\eea

This is a novel class of algebras with fractional derivatives which we call F-Galilei algebra.
Similar to L-Galilei algebra given any positive integer value for $F$ we a have a new algebra. Interestingly similar to the L-Galilei algebra the first element of the class has dynamical index $z=2\;\;\;\;(F=1)$ but differs from the Schr\"odinger algebra by $X^1$. The second element of   F-Galilei class has dynamical index  $z=\frac{3}{2}$. This is the algebra obtained by Henkel in \cite{Henkelpheno}. It can be deduced from  the weights diagram that $Q_F$ commutes with all the elements of the algebra, i.e. it is  the central charge.  
  
\begin{figure}
\includegraphics[width=.8\textwidth]{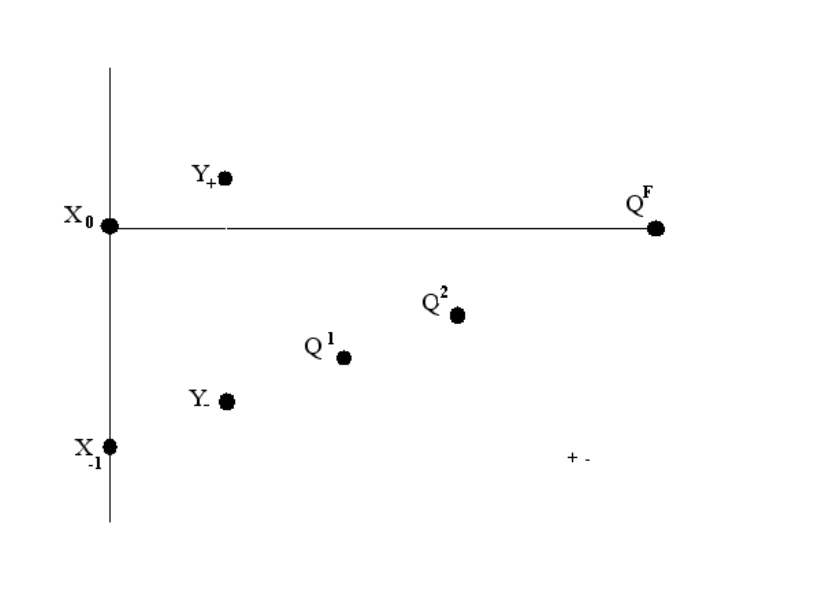}
\caption{Weights of F-Galilei algebra. Vertical axis shows the scaling weight and $Q_F$ is the central charge.}\label{fig}
\end{figure}


\section{Conclusions}

We have observed that admitting fractional derivatives greatly enlarges the kind of symmetries we can have. In this paper we added fractional derivative generators to the scale, space and time invariance, results in a class of algebras with an infinite number of elements. The first element of this class is a sub-algebra of the Schr\"odinger algebra where only $X^1=-t^2\partial_t-2tx\partial_x$ is left out. This means that the algebra has scale invariance within it but not conformal invariance. Note that for $F=1$ we do not have fractional derivatives. 

The second member of the class is an algebra which is has dynamical index $z=3/2$ and previously was found in \cite{Henkelpheno} . We have not yet succeeded to find candidate actions left invariant by this algebras, which we hope to do in future. However it is more than likely that these will be non-local actions.

\section*{Acknowledgments}

We aknowledge M. Henkel for discussion.



\end{document}